\documentclass[10pt, conference, letterpaper]{IEEEtran}
\usepackage{url}
\usepackage[utf8]{inputenc}
\usepackage[greek,english]{babel}
\usepackage{xcolor}
\usepackage{color,soul}
\usepackage{amsmath}
\usepackage{amssymb}
\usepackage{microtype}
\usepackage{textcomp}

\usepackage{graphicx}
\usepackage{tablefootnote}
\usepackage{booktabs}
\usepackage{tabularx}
\usepackage{tikz}
\usepackage{pgfplots}
\pgfplotsset{compat=newest}
\pgfplotsset{plot coordinates/math parser=false}
\newlength\fheight
\newlength\fwidth
\usetikzlibrary{plotmarks,patterns,decorations.pathreplacing,backgrounds,calc,
  arrows,arrows.meta,spy,matrix}
\usepgfplotslibrary{patchplots,groupplots}
\usepackage{tikzscale}

\usepackage{multirow}
\usepackage{tkz-kiviat}
\usepackage[htt]{hyphenat}

\usepackage[font=footnotesize]{subcaption}
\usepackage[font=footnotesize]{caption}

\usepackage{mathtools}

\usepackage{dblfloatfix}    
\usepackage{colortbl}

\usepackage{makecell}
\usepackage{diagbox}
\usepackage{tikz-qtree}
\usetikzlibrary{trees} 

\usepackage{cite}
\usepackage{bbold}
\usepackage{bbm}

\usepackage{array}
\newcolumntype{?}{!{\vrule width 1.5pt}}
\usepackage{makecell}
\usepackage{mdframed}
\usepackage[many]{tcolorbox}
\usepackage{empheq}

\newtcbox{\mybox}[1][]{nobeforeafter,math upper,tcbox raise base,
  enhanced,frame hidden,boxrule=0pt,interior style={top color=green!10!white,
  bottom color=green!10!white,middle color=green!50!yellow},
  fuzzy halo=1pt with green,drop large lifted shadow,#1}

\usepackage{siunitx}
\usepackage{fancyhdr}
\fancypagestyle{FirstPage}{
  \chead{This work has been submitted to the IEEE for possible publication.
    Copyright may be transferred without notice,
    after which this version may no longer be accessible.}
}

\usepackage[acronyms,nonumberlist,nopostdot,nomain,nogroupskip]{glossaries}

\newacronym{vanet}{VANET}{Vehicular Ad Hoc Network}
\newacronym{ukf}{UKF}{Unscented Kalman Filter}
\newacronym{cdf}{CDF}{Cumulative Distribution Function}
\newacronym{iid}{IID}{Independent and Identically Distributed}
\newacronym{ctra}{CTRA}{Constant Turn Rate and Acceleration}
\newacronym{kf}{KF}{Kalman Filter}
\newacronym{ekf}{EKF}{Extended Kalman Filter}
\newacronym{pf}{PF}{Particle Filter}
\newacronym{hmm}{HMM}{Hidden Markov Model}
\newacronym{its}{ITS}{Intelligent Transport System}
\newacronym{ml}{ML}{Machine Learning}
\newacronym{svm}{SVM}{Support Vector Machine}
\newacronym{nn}{NN}{Neural Network}
\newacronym{gps}{GPS}{Global Positioning System}
\newacronym{dr}{DR}{Dead Reckoning}
\newacronym{bf}{BF}{Bayesian Filtering}
\newacronym{rbf}{RBF}{Radial Basis Function}
\newacronym{c-its}{C-ITS}{Connected and Intelligent Transportation System}
\newacronym{sumo}{SUMO}{Simulation of Urban MObility}
\newacronym{dsrc}{DSRC}{Dedicated Short Range Communication}
\newacronym{csmaca}{CSMA/CA}{Carrier Sense Multiple Access with Collision Avoidance}
\newacronym{qoi}{QoI}{Quality of Information}
\newacronym{aoi}{AoI}{Age of Information}
\newacronym{mac}{MAC}{Medium Access Control}
\newacronym{pdf}{PDF}{Probability Density Function}
\newacronym{uav}{UAV}{Unmanned Aerial Vehicle}
\newacronym{3dctra}{3D-CTRA}{3-Dimensional CTRA}
\newacronym{lora}{LoRaWAN}{Long Range Wide Area Network}
\newacronym{lpwan}{LPWAN}{Low Power Wide Area Network}
\newacronym{dc}{DC}{Duty Cycle}
\newacronym{ul}{UL}{Uplink}
\newacronym{ns}{NS}{Network Server}
\newacronym{gw}{GW}{Gateway}
\newacronym{ed}{ED}{End Device}
\newacronym{sf}{SF}{Spreading Factor}

\begin{document}
\title{Combining LoRaWAN and a New 3D Motion Model for Remote UAV Tracking}

\author{\IEEEauthorblockN{Federico Mason, Federico Chiariotti, Martina Capuzzo, Davide Magrin, Andrea Zanella, Michele Zorzi}
  \IEEEauthorblockA{Department of Information Engineering, University of Padova -- Via Gradenigo, 6/b, 35131 Padova, Italy\\
    Email: {\tt\footnotesize\{masonfed, chiariot, capuzzom, magrinda, zanella,
      zorzi\}@dei.unipd.it} }}
\maketitle

\thispagestyle{FirstPage}

\begin{abstract}
  Over the last few years, the many uses of \glspl{uav} have captured the
  interest of both the scientific and the industrial communities. A typical
  scenario consists in the use of \glspl{uav} for surveillance or target-search
  missions over a wide geographical area. In this case, it is fundamental for
  the command center to accurately estimate and track the trajectories of the
  \glspl{uav} by exploiting their periodic state reports. In this work, we
  design an ad hoc tracking system that exploits the \gls{lora} standard for
  communication and an extended version of the \gls{ctra} motion model to
  predict drone movements in a 3D environment. Simulation results on a publicly
  available dataset show that our system can reliably estimate the position and
  trajectory of a \gls{uav}, significantly outperforming baseline tracking
  approaches.\end{abstract}

\glsresetall

\section{Introduction}\label{sec:intro}
Over the last few years, \glspl{uav} have entered the mainstream: the
commercialization of low-cost drones for amateur and professional use is quickly
increasing the number of flying units, which will soon be measured in millions,
according to the U.S. Federal Aviation Administration (FAA)\footnote{FAA
  Aerospace Forecast, Fiscal Years 2019-2039:
  \url{https://www.faa.gov/news/updates/?newsId=93646}}. Their integration in
cellular networks, both as end-users and as coverage
extenders~\cite{fotouhi2019survey}, is already being discussed, and 5G systems
are expected to make use of \glspl{uav} of different sizes, from small-scale
low-altitude drones to communication satellites~\cite{sekander2018multi}.
Although energy and battery concerns are still critical~\cite{long2018energy},
the use of \glspl{uav} is being proposed for several kinds of scenarios, from
remote infrastructure monitoring~\cite{hament2018unmanned} to disaster
monitoring~\cite{zakaria2018aerial} and relief~\cite{scherer2015autonomous}.

As the capabilities of \glspl{uav} evolve towards the full support of
safety-critical applications, accurate positioning of drones is going to become
more and more important. Although \glspl{uav} often have on-board \gls{gps}
receivers, filtering~\cite{dentler2016real} and data fusion techniques, often
integrating camera image processing~\cite{angelino2012uav}, can significantly
improve the positioning accuracy by combining several measurements into a single
solution that is more robust and precise than any individual approach.

In this work, we propose a system to remotely track the position of a \gls{uav}
moving in a 3D environment. In the considered scenario, a mission control
station exploits a novel 3D motion model, called \gls{3dctra}, to follow the
trajectory of the target. Our model extends the well-known \gls{ctra} model,
widely used in vehicular scenarios, by adding a third dimension which allows it
to represent even complex banking maneuvers accurately. We also study a simpler
model, named \gls{ctra}+, which considers linear motion on the vertical axis. In
both cases, the tracking mechanism is the same: the \gls{uav} periodically
transmits its state, including the heading, speed and acceleration, and the
control station estimates the target position by evolving the motion model. In
this way, even sporadic updates allows the system to accurately track the
\gls{uav}.

In our system, state updates are transmitted through the \gls{lora}
communication standard. This technology allows the transmission of low-bitrate
messages at very long distances, enabling the control station to track the drone
at ranges of several kilometers with minimal infrastructure. Considering the
limited duty cycle imposed by the \gls{lora} specifications, our system can
support swarms of up to 72 drones with a packet collision rate below 10\% by
using different \glspl{sf}~\cite{haxhibeqiri2017lora}. When \gls{lora} is tuned
to achieve larger communication range, the intervals between transmissions can
last several seconds, thus making the tracking more difficult. It is hence
interesting to analyze the feasibility of such a framework, and to investigate
its performance when varying the considered mobility model. We tested our system
in extensive ns-3 simulations using the \gls{uav} mobility traces from the
Mid-Air public dataset and comparing the two mobility models we proposed against
a baseline solution implementing \gls{dr}, a well-known tracking method
exploiting a uniform rectilinear motion model to predict the target movements.
The results show that the more accurate \gls{3dctra} mobility model can bring an
improvement of up to 30\% on the 75th percentile tracking error.

The rest of the paper is organized as follows. Sec.~\ref{sec:related} presents
the state of the art on \gls{uav} applications and tracking models, including
both \gls{gps} and visual data. Sec.~\ref{sec:model} presents the \gls{ctra}+
and \gls{3dctra} models, including the relative update equations, and describes
the \gls{lora} standard and the frequency plan needed for our application. The
simulation settings and the results are described in Sec.~\ref{sec:results},
while Sec.~\ref{sec:conclusion} presents our concluding remarks and ideas for
future work.

\section{Related work}\label{sec:related}
\glspl{uav}' popularity has grown exponentially over the past few years, and
their widespread use could enable a real Internet of Flying
Robots~\cite{huang2018towards} in the near future. Drones are used for
environmental monitoring in a wide range of scenarios, from traffic jam
detection~\cite{sharma2018lorawan} to industry and
agriculture~\cite{vasisht2017farmbeats}, and are posed to become a key Smart
City infrastructure~\cite{ismail2018internet}. \glspl{uav} are also being used
in combination with ground-based robots to help them perform complex
tasks~\cite{arbanas2018decentralized}. However, disaster management and relief
is perhaps the most interesting application for \glspl{uav}: drones can easily
avoid ground-level obstacles and flooded areas by flying over them, surveying
the extent of the damage~\cite{zakaria2018aerial} or helping with search and
rescue operations~\cite{scherer2015autonomous} and communications. In order to
enable these critical services, controllers must be able to follow and even
anticipate the drone's trajectory. This requires the \gls{uav} to transmit
frequent positioning updates~\cite{jawhar2017communication}, often at long
ranges.

The target tracking problem is a well-studied research topic, and is usually
solved by representing the target's motion using simple models and estimating
its position with a \gls{bf} algorithm. The best-known \gls{bf} algorithms used
in this context are the \gls{kf} \cite{Kalman:1960} and the \gls{pf}
\cite{Moral:1996}. Long-term forecasting can be achieved by simply applying the
predictive step of the \gls{bf} to the last available state estimation. However,
this solution does not provide good performance when updates are infrequent,
especially if the model is inaccurate. In this perspective, our work tries to
minimize broadcasting operations while ensuring accurate position estimation.

The tracking problem has been widely explored in 2D vehicular
scenarios~\cite{mason2019quality}, often using the \gls{ctra}
model~\cite{Tsogas:2005}, which considers an accelerating vehicle with constant
turn rate. A similar model for drones moving horizontally was presented
in~\cite{biomo2014enhanced}, including Gaussian noise on the motion parameters.
A more complex model with several possible maneuvers was described
in~\cite{bouachir2014mobility}, adapting the \gls{ctra} settings to draw the
correct trajectory. In general, motion models for drones are based on 2D
\gls{ctra} or simpler models with constant speed~\cite{wan2013smooth} or
heading~\cite{tiemann2015design}. To the best of our knowledge, our \gls{ctra}+
and \gls{3dctra} models are the first models that can represent 3D maneuvers
with the same flexibility that \gls{ctra} has in the 2D space.

\section{System model}\label{sec:model}

In this work, we model a \gls{uav} which periodically transmits its state to a
control station using the \gls{lora} communication standard. The aim of the
control station is to accurately track the \gls{uav} position in different
scenarios. To represent the drone motion in a 3D environment, we consider three
possible models, i.e., \gls{dr}, \gls{ctra}+, and \gls{3dctra}. In the rest of
the section, we first extend conventional \gls{ctra}, obtaining the system
equations for \gls{ctra}+ and \gls{3dctra}. Then, we analyze the tracking and
communication frameworks.

\subsection{The CTRA+ model}

While standard \gls{ctra} only tracks the yaw, i.e., the angle $\theta$ between
the drone's heading and the reference direction on the horizontal plane, 3D
motion models must also consider the pitch, i.e., the vertical angle $\phi$
between the drone's heading and the horizon. Moreover, the target state must
include the altitude $z$ as well as the horizontal position $(x,y)$, resulting
in the 5-tuple $(x, y, z, \theta, \phi)$. These parameters are common to all the
motion models we implement. However, none of our models explicitly considers
roll, which is not strictly necessary to represent motion in a 3D space.

In 2D \gls{ctra}, the turn rate $\omega=\frac{d\theta}{dt}$ is assumed to be
constant. The \gls{ctra}+ model makes the same assumption and, moreover,
considers a constant pitch $\phi$:
\begin{align}
  \theta(t)&=\theta(0)+\omega t\label{eq:theta}\\
  \phi(t)&=\phi(0)\label{eq:phi_plus},
\end{align}
where $\theta(0)$ and $\phi(0)$ represent the initial heading of the drone.

Like standard \gls{ctra}, \gls{ctra}+ assumes the tangential acceleration
$a=\frac{dv}{dt}$ to be constant, which turns the circular motions into
Archimedean spirals~\cite{archimedes225spirals}. In particular, \gls{ctra}+
considers the spirals on a plane tilted by an angle $\phi$ with respect to the
horizon. By projecting the \gls{uav}'s velocity vector $\mathbf{v}(t)$, we can
get its three components:
\begin{align}
  v_x(t)=\frac{dx}{dt}&=v(t)\cos(\theta(t))\cos(\phi(t));\\
  v_y(t)=\frac{dy}{dt}&=v(t)\sin(\theta(t))\cos(\phi(t));\\
  v_z(t)=\frac{dz}{dt}&=v(t)\sin(\phi(t)).
\end{align}
Therefore, the velocity's magnitude $v(t)$ is given by:
\begin{align}
  v(t)&=\left(v_x(t)\right)^2+\left(v_y(t)\right)^2+\left(v_z(t)\right)^2.
\end{align}
In order to compute the position at any time, we need to integrate the velocity
components over time:
\begin{align}
  x(t)&=x(0)+\int_0^t v(\tau)\cos(\theta(\tau))\cos(\phi)d\tau;\\
  y(t)&=y(0)+\int_0^t v(\tau)\sin(\theta(\tau))\cos(\phi)d\tau;\\
  z(t)&=z(0)+\int_0^t v(\tau)\sin(\phi)d\tau.
\end{align}
We note that the procedure is equivalent to 2D \gls{ctra}~\cite{Tsogas:2005} for
the $x$ and $y$ components, except for the constant multiplying factor
$\sin(\phi)$. Hence, the \gls{ctra}+ state is given by:
\begin{align}
  \mathbf{x}_{\text{CTRA+}}(t)&=\left[x(t)\ y(t)\ z(t)\ \theta(t)\ \phi\ v(t)\ a\ \omega \right]^T,\label{eq:plus_state}
\end{align}
which corresponds to the tuple representing the current attitude, with the
addition of the velocity $v$, the acceleration $a$, and the turn rate $\omega$.

\subsection{The 3D-CTRA model}

The \gls{3dctra} model extends the above description by adding a constant tilt
rate $\psi=\frac{d\phi}{dt}$. Consequently, the \gls{uav}'s movement is
represented as the combination of two independent spiraling motions on the
horizontal and vertical planes, forming a curved helix. While the evolution of
$\theta(t)$ still follows~\eqref{eq:theta}, the pitch is given by:
\begin{align}
  \phi(t)&=\phi(0)+\psi t\label{eq:phi_3d}.
\end{align}
This complicates the derivation of the motion equations considerably, since
$\phi(t)$ is now time-dependent. For the sake of simplicity, we report the
procedure only for $x(t)$, which is given by
\begin{align}
  x(t)&=x(0)+\int_0^t v(\tau)\cos(\theta(\tau))\cos(\phi(\tau))d\tau.
\end{align}
Applying the Werner formula, we obtain
\begin{equation}
  \begin{split}
		x(t)=x(0)+\int_0^t \frac{v}{2} ( & \cos(\theta(\tau)+\phi(\tau))\\ + &
    \cos(\theta(\tau)-\phi(\tau)) ) d\tau,
  \end{split}
\end{equation}
which can be solved in closed form. The derivations for $y(t)$ and $z(t)$ follow
the same steps: the final results are given
in~\eqref{eq:x_int}-\eqref{eq:z_int}, on top of next page, where we used the
compact notation $[F(x)]_a^b=F(b)-F(a)$ to indicate that the primitive function
$F(x)$ should be evaluated at the extremes $a$ and $b$.
\begin{table*}[t]
  \begin{align}
    x(t)&=x(0)+\left[\frac{v(\tau)\sin(\theta(\tau)+\phi(\tau))}{2(\omega+\psi)}+\frac{v(\tau)\sin(\theta(\tau)-\phi(\tau))}{2(\omega-\psi)}+\frac{a\cos(\theta(\tau)+\phi(\tau))}{2(\omega+\psi)^2}+\frac{a\cos(\theta(\tau)-\phi(\tau))}{2(\omega-\psi)^2}\right]_0^t\label{eq:x_int}\\
    y(t)&=y(0)+\left[-\frac{v(\tau)\cos(\theta(\tau)+\phi(\tau))}{2(\omega+\psi)}-\frac{v(\tau)\cos(\theta(\tau)-\phi(\tau))}{2(\omega-\psi)}+\frac{a\sin(\theta(\tau)+\phi(\tau))}{2(\omega+\psi)^2}+\frac{a\sin(\theta(\tau)-\phi(\tau))}{2(\omega-\psi)^2}\right]_0^t\label{eq:y_int}\\
    z(t)&=z(0)+\left[-v(\tau)\frac{\cos(\phi(\tau))}{\psi}+a\frac{\sin(\phi(\tau))}{\psi^2}\right]_0^t.\label{eq:z_int}
  \end{align}
  \vspace{-0.5cm}
\end{table*}

Finally, three special cases need to be considered. First, when $\psi=0$, i.e.,
when the model is equivalent to \gls{ctra}+ and the pitch is constant, the value
of $z(t)$ becomes:
\begin{align}
  z(t)&=z(0)+\sin(\phi(t))\left(v(t)t-\frac{at^2}{2}\right).
\end{align}
Then, when $\omega=\psi$, i.e., the rotations on the two axes have the same
period, the values of $x(t)$ and $y(t)$ become:
\begin{align}
  &\begin{split}
    x(t)= \Bigg[&\frac{v(\tau)\sin(\theta(\tau)+\phi(\tau))}{2(\omega+\psi)}+\frac{a\cos(\theta(\tau)+\phi(\tau))}{2(\omega+\psi)^2}+\\
    &\left(\frac{v(\tau)\tau}{2}-\frac{a\tau^2}{4}\right)\cos(\theta(\tau)-\phi(\tau))\Bigg]_0^t+x(0)\label{eq:x_spec1}
  \end{split}\\
  &\begin{split}
    y(t)= \Bigg[&-\frac{v(\tau)\cos(\theta(\tau)+\phi(\tau))}{2(\omega+\psi)}+\frac{a\sin(\theta(\tau)+\phi(\tau))}{2(\omega+\psi)^2}+\\
    &\left(\frac{v(\tau)\tau}{2}-\frac{a\tau^2}{4}\right)\sin(\theta(\tau)-\phi(\tau))\Bigg]_0^t+y(0)\label{eq:y_spec1}
  \end{split}
\end{align}
The case in which $\omega=-\psi$ produces a similar result, with inverted terms.
Setting $t=T$,~\eqref{eq:theta},~\eqref{eq:phi_3d}
and~\eqref{eq:x_int}-\eqref{eq:z_int}, or their special case equivalents, define
the full non-linear version of the \gls{3dctra} model with step $T$. In
particular, the \gls{3dctra} state is given by:
\begin{align}
  \mathbf{x}_{\text{3D-CTRA}}(t)&=\left[x(t)\ y(t)\ z(t)\ \theta(t)\ \phi(t)\ v(t)\ a\ \omega\ \psi \right]^T,\label{eq:3d_state}
\end{align}
which is equivalent to~\eqref{eq:plus_state}, with the addition of the tilt rate
$\psi$.

We observe that \gls{3dctra} considers constant values for both $\omega$ and
$\psi$. This does not reflect the real behavior of an aircraft, as dives and
climbs are usually relatively short. To overcome this problem and make the model
more realistic, we make the tracking system reduce the value of $\psi$ by a
factor $\eta$ after every prediction step. In other words, the model assumes
that the drone will gradually reduce its tilt rate and stabilize its pitch.

\subsection{Remote tracking and communications}
As in~\cite{Tsogas:2005}, the tracking process is implemented by the \gls{ukf}
algorithm. In particular, we assume that the \gls{uav} and the control station
are equipped with two \gls{ukf}s~\cite{van2004sigma}. While the drone exploits
the measurements provided by its on-board sensors to track its own state, the
control station's \gls{ukf} has no input but the information received from the
\gls{uav}. We adopt a periodic broadcasting strategy~\cite{mason2019quality}:
the \gls{uav} sends the estimate of its own state to the control station with a
constant inter-transmission period. After it receives an update, the control
station updates its \gls{ukf} with the new information and exploits the
predictive step to forecast the \gls{uav}'s trajectory. Naturally, the errors
will compound, causing long-term predictions to become less and less accurate
until the next update.

In order to enable the \gls{uav} to send the \gls{ukf} parameters even at great
distances, we considered the \gls{lora} technology~\cite{lorawan}, which
leverages the proprietary LoRa PHY modulation which is based on a chirp spread
spectrum technique to transmit over long distances. The performance of the
modulation can be tuned through the \gls{sf} parameter, which takes values from
7 to 12, and allows to trade coverage range for data rate: signals transmitted
with higher \glspl{sf} values require longer transmission times, but are more
robust to channel impairments and, thus, can be received at farther distances,
up to several kilometers in open-air scenarios.

\gls{lora} is based on a star topology, with three kinds of devices:
\begin{itemize}
\item the \textit{\gls{ns}}, which is the central network controller;
\item the \textit{\glspl{ed}}, peripheral nodes, collecting data and
  transmitting them through the LoRa modulation;
\item the \textit{\glspl{gw}}, acting as relays between \glspl{ed} and \gls{ns},
  collecting messages from devices and forwarding them to the controller through
  a legacy IP connection, and \textit{vice-versa}.
\end{itemize}
We assume that the drone is equipped with a \gls{lora} \textit{Class A}
\gls{ed}. This class is designed to consume a minimum amount of energy, staying
in sleep mode most of the time, transmitting when necessary, and opening two
short windows for reception after each transmission.
\gls{lora} works in the unlicensed 868~MHz sub-band, which is subject to
\gls{dc} regulations. In particular, three 125~MHz channels are allocated to
\gls{ul} transmissions, and must respect a \gls{dc} limitation of 1\%. Another
option is to use a single 250~MHz channel, which does not bring the benefits of
frequency orthogonality but reduces the packet transmission time, and is
preferable in case of a system with a single drone.

\begin{figure}[!t] \setlength{\belowcaptionskip}{-0.5cm} \centering
  \resizebox{\columnwidth}{!}{ \large{ \tikzset{ block/.style = {draw,
          rectangle, minimum height = 3cm, minimum width = 1em}}
      \begin{tikzpicture}[auto]
        \node at (-1, 3)[name=dr]{DR}; \node at (-1, 1.5)[name=plus]{CTRA+};
        \node at (-1, 0)[name=3d]{3D-CTRA}; \node at (4, 4)[name=4b]{4 B}; \node
        at (8, 4)[name=8b]{8 B}; \node at (12, 4)[name=12b]{12 B}; \node at (4,
        0)[name=4bb]{}; \node at (8, 0)[name=8bb]{}; \node at (12,
        0)[name=12bb]{}; \node at (1,3)[rectangle,draw,inner sep=0pt,minimum
        width=2cm,minimum height=1cm,fill={blue!20},name=xdr] {$x(t)$};
        \node[rectangle, right of=xdr,node distance=2cm,draw,inner
        sep=0pt,minimum width=2cm,fill={blue!20},minimum height=1cm,name=ydr]
        {$y(t)$}; \node[rectangle, right of=ydr,node distance=2cm,draw,inner
        sep=0pt,minimum width=2cm,minimum height=1cm,fill={blue!20},name=zdr]
        {$z(t)$}; \node[rectangle, right of=zdr,node distance=1.5cm,draw,inner
        sep=0pt,minimum width=1cm,fill={blue!20},minimum
        height=1cm,name=thetadr] {$\theta$}; \node[rectangle, right
        of=thetadr,node distance=1cm,draw,inner sep=0pt,minimum
        width=1cm,fill={blue!20},minimum height=1cm,name=phidr] {$\phi$};
        \node[rectangle, right of=phidr,node distance=1cm,draw,inner
        sep=0pt,minimum width=1cm,fill={blue!20},minimum height=1cm,name=vdr]
        {$v(t)$}; \node at (1,1.5)[rectangle,draw,inner sep=0pt,minimum
        width=2cm,minimum height=1cm,fill={blue!20},name=xplus] {$x(t)$};
        \node[rectangle, right of=xplus,node distance=2cm,draw,inner
        sep=0pt,minimum width=2cm,fill={blue!20},minimum height=1cm,name=yplus]
        {$y(t)$}; \node[rectangle, right of=yplus,node distance=2cm,draw,inner
        sep=0pt,minimum width=2cm,minimum height=1cm,fill={blue!20},name=zplus]
        {$z(t)$}; \node[rectangle, right of=zplus,node distance=1.5cm,draw,inner
        sep=0pt,minimum width=1cm,fill={blue!20},minimum
        height=1cm,name=thetaplus] {$\theta(t)$}; \node[rectangle, right
        of=thetaplus,node distance=1cm,draw,inner sep=0pt,minimum
        width=1cm,fill={blue!20},minimum height=1cm,name=phiplus] {$\phi$};
        \node[rectangle, right of=phiplus,node distance=1cm,draw,inner
        sep=0pt,minimum width=1cm,fill={blue!20},minimum height=1cm,name=vplus]
        {$v(t)$}; \node[rectangle, right of=vplus,node distance=1cm,draw,inner
        sep=0pt,minimum width=1cm,fill={green!20},minimum height=1cm,name=aplus]
        {$a$}; \node[rectangle, right of=aplus,node distance=1cm,draw,inner
        sep=0pt,minimum width=1cm,fill={green!20},minimum
        height=1cm,name=omegaplus] {$\omega$}; \node at
        (1,0)[rectangle,draw,inner sep=0pt,minimum width=2cm,minimum
        height=1cm,fill={blue!20},name=x3d] {$x(t)$}; \node[rectangle, right
        of=x3d,node distance=2cm,draw,inner sep=0pt,minimum
        width=2cm,fill={blue!20},minimum height=1cm,name=y3d] {$y(t)$};
        \node[rectangle, right of=y3d,node distance=2cm,draw,inner
        sep=0pt,minimum width=2cm,minimum height=1cm,fill={blue!20},name=z3d]
        {$z(t)$}; \node[rectangle, right of=z3d,node distance=1.5cm,draw,inner
        sep=0pt,minimum width=1cm,fill={blue!20},minimum
        height=1cm,name=theta3d] {$\theta(t)$}; \node[rectangle, right
        of=theta3d,node distance=1cm,draw,inner sep=0pt,minimum
        width=1cm,fill={blue!20},minimum height=1cm,name=phi3d] {$\phi(t)$};
        \node[rectangle, right of=phi3d,node distance=1cm,draw,inner
        sep=0pt,minimum width=1cm,fill={blue!20},minimum height=1cm,name=v3d]
        {$v(t)$}; \node[rectangle, right of=v3d,node distance=1cm,draw,inner
        sep=0pt,minimum width=1cm,fill={green!20},minimum height=1cm,name=a3d]
        {$a$}; \node[rectangle, right of=a3d,node distance=1cm,draw,inner
        sep=0pt,minimum width=1cm,fill={green!20},minimum
        height=1cm,name=omega3d] {$\omega$}; \node[rectangle, right
        of=omega3d,node distance=1cm,draw,inner sep=0pt,minimum
        width=1cm,fill={red!20},minimum height=1cm,name=psi3d] {$\psi$};

        \draw[dashed] (4b.south) to (4bb.north); \draw[dashed] (8b.south) to
        (8bb.north); \draw[dashed] (12b.south) to (12bb.north);

\end{tikzpicture}
}}
\caption{Schematic of the payload format for the three tracking schemes}
\label{fig:payload}
\end{figure}
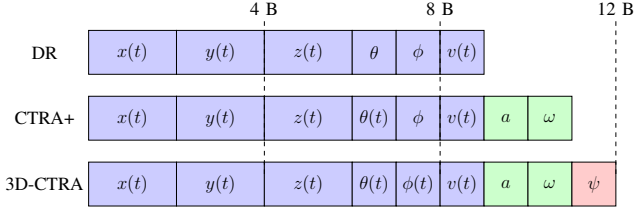

Because of the \gls{dc} limitation, we need to compress the system state to
reduce the inter-transmission time and improve the tracking performance. In
order to minimize the payload size, we can represent the position using 2 bytes,
allowing movement in a square box with a size of 13~km while limiting the
quantization error to 10~cm, significantly less than the average \gls{gps}
error. Angles and turn rates can be represented using just 1 byte, with a
maximum error of 0.7 degrees; since velocity and acceleration are limited, they
can also be represented with just 1 byte, with a negligible loss of precision.
Since \gls{dr} tracking requires the knowledge of the attitude 5-tuple and the
velocity, its minimum payload size is 9 bytes. The \gls{ctra}+ state as given
in~\eqref{eq:plus_state} requires 11~bytes, and the \gls{3dctra} state as given
in~\eqref{eq:3d_state} requires 12 bytes. The different payload formats are
reported in Fig.~\ref{fig:payload}. The LoRa transmission times for packets with
these lengths are reported in Tab.~\ref{tab:freq_plan}: to respect the \gls{dc}
limitation, packets can be sent only sporadically, with a transmission period in
the order of a few seconds.

\begin{table}[b]
  \vspace{-0.5cm} \centering
	\begin{tabularx}{1.0\linewidth}[c]{c|cccc}
		\toprule
		SF & B (MHz) &
                   \makecell{Packet size (B)} &
                                                \makecell{Transmission\\ time (s)} &
                                                                                     \makecell{Min transmission \\ interval (s)} \\
		\midrule
		\multirow{2}{*}{7} & 125 & 9, 11, 12 & 0.0412 & 4.21 \\
    \cmidrule{2-5}
       &250 & 9, 11, 12 & 0.0206 & 2.06 \\
		\midrule
		\multirow{4}{*}{8} & \multirow{2}{*}{125} & 9 & 0.0722 & 7.22 \\
       && 11, 12 & 0.0824 & 8.24 \\
    \cmidrule{2-5}
       &\multirow{2}{*}{250} & 9 & 0.0391 & 3.61 \\
       && 11, 12 & 0.0412 & 4.12 \\
		\bottomrule
	\end{tabularx}
	\caption{Transmission times for packets with the three different payloads and
    minimum transmission interval required to respect \gls{dc} regulations with
    the standard frequency plan.}
	\label{tab:freq_plan}
\end{table}


\section{Simulation settings and results}\label{sec:results}
To simulate the scenario described in the previous section, we exploit the
Mid-Air dataset~\cite{Fonder2019MidAir}, which contains the flying records of a
quad-copter moving in 3 different virtual environments. In particular, we
consider 30 trajectories for a total flying duration of 44 minutes. These data
are used to represent the ground-truth motion of the target while we
synthetically generated noisy data to represent the information acquired with
the drone's sensors. The sensor data included the position, the attitude, and
the velocity and acceleration vectors of the \gls{uav}, combining \gls{gps},
accelerometer and gyroscope. In the rest of the section, we describe the setting
of our simulations and we present the obtained results.

\subsection{Settings}

In our scenario, the control station is located at a distance $d$ from the
drone. The process noise of the tracking system is described by the matrix $Q =
q I$, where $I$ represents the identity matrix. Instead, the error affecting the
drone measurements is given by a diagonal matrix $R$, whose elements represent
the accuracy of the various drone sensors. The noise matrices and the \gls{ukf}
parameters are reported in Tab.~\ref{tab:track_param}. In particular, values of
$R$ were chosen according to \cite{team2014global, kim2014probabilistic,
  falco2017loose}. We highlight that the \gls{ukf} setting, e.g., the state
dimension, changes according to the chosen motion model. As already stated, the
\gls{ukf} at the control station is used to estimate the target trajectory by
exploiting only the predictive step. This implies that, when a new update is
received, the filter state is substituted with the new information, and the
estimation process starts again.

\begin{table}[b]
  \vspace{-0.5cm} \centering
	\begin{tabularx}{1.0\linewidth}[c]{ccc}
		\toprule
		Parameter & Value &	Description \\
		\midrule
		$R_{x}$ & 0.8274 $m^2$ & Position accuracy along x \\
		$R_{y}$ & 0.8274 $m^2$ & Position accuracy along y \\
		$R_{z}$ & 3.7481 $m^2$ & Position accuracy along z \\
		$R_{v}$ & 0.2500 $(m/s)^2$ & Speed accuracy \\
		$R_{a}$ & 0.1521 $(m/s^2)^2$ & Acceleration accuracy \\
		$R_{\theta}$ & 0.0085 $rad^2$ & Yaw accuracy \\
		$R_{\phi}$ & 0.0085 $rad^2$ & Pitch accuracy \\
		$R_{\omega}$ & 0.0003 $(rad/s)^2$ & Turn rate accuracy\\
		$R_{\psi}$ & 0.0003 $(rad/s)^2$ & Tilt rate accuracy\\
		$q$ & 0.1 & Process noise matrix parameter\\
		$\eta$ & 0.9 & Tilt reduction parameter\\
		\bottomrule
	\end{tabularx}
	\caption{Tracking system parameters.}
	\label{tab:track_param}
\end{table}

The scenario of interest was studied with the network simulator ns-3, with a
single drone moving in the space according to the mobility traces
of~\cite{majdik2017zurich}. The drone was equipped with a \gls{lora} interface,
which transmitted packets at the maximum frequency allowed by the \gls{dc}.
These messages were collected by a \gls{gw}, and forwarded to a \gls{ns}.
Transmitted packets did not require any acknowledgment, and the \gls{ns} did not
control any of the communication parameters. For each packet, we recorded
whether it was successfully received or not, and estimate the tracking
performance. We also moved the initial position of the \gls{gw} to see how much
the tracking performance was affected by the communication limitations. In the
rest of the section, we will analyze the positioning error for different
tracking and communication scenarios. In particular, we investigate our tracking
scheme for different values of the \gls{sf} and of the initial distance $d$
between the \gls{uav} and the \gls{gw}.

\subsection{Results}
\begin{figure}[t!]
	\centering \setlength{\belowcaptionskip}{-0.5cm}
  \includegraphics[width=0.9\columnwidth,trim={0cm 1.2cm 0cm
    1cm},clip]{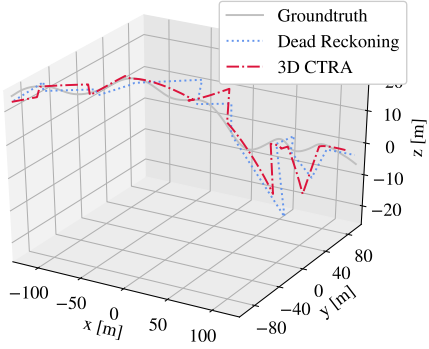}
	\caption{3D trajectory of the UAV.}
	\label{fig:trajectory}
\end{figure}

\begin{figure}[t!]
	\centering \setlength{\belowcaptionskip}{-0.5cm}
  \includegraphics[width=0.7\columnwidth,trim={0.5cm 0 1.5cm
    1cm},clip]{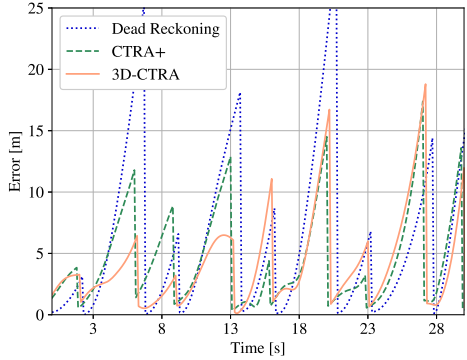}
	\caption{Error over time with $d=1000$ m, SF = 7 and $B=250$ MHz.}
	\label{fig:err}
\end{figure}

First, we consider the 30 s drone trajectory shown in Fig.~\ref{fig:trajectory}.
The same figure includes the trajectories estimated by the control station using
the \gls{dr} and \gls{3dctra} motion models, considering a communication setup
with $d=1000$ m, \gls{sf} = 7 and $B=250$ MHz. Comparing the different
trajectories, we observe how the \gls{dr} scheme does not follow the target
while the \gls{3dctra} scheme has smaller deviations from the real trajectories.
This is confirmed by the results in Fig.~\ref{fig:err}, which shows the base
station tracking error over time for all the considered models. We highlight
that the error of the \gls{dr} model rapidly increases every time the drone
performs non-linear movements, as it happens at time $t\simeq6,12,19$ s.
Instead, the error of both the \gls{ctra}+ and the \gls{3dctra} models presents
a smoother trend, with fewer and lower peaks.

\begin{figure}[t!]
	\centering \setlength{\belowcaptionskip}{-0.5cm}
  \includegraphics[width=0.7\columnwidth,trim={0.5cm 0 1.5cm
    1cm},clip]{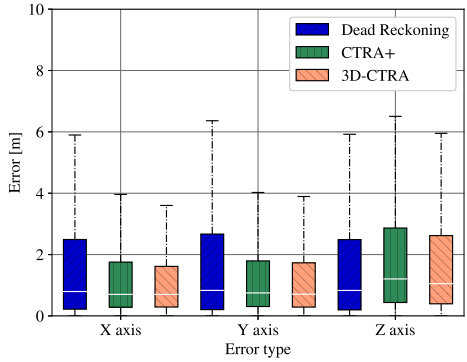}
	\caption{Error over the three axes with $d=1000$ m with SF = 7 and $B=250$
    MHz.}
	\label{fig:err_dr6_d1000}
\end{figure}

From here on, we consider the cumulative results over all the available
trajectories. Fig. \ref{fig:err_dr6_d1000} shows the boxplot of the position
error along the three axes with SF = 7, $d=1000$ m, and $B=250$ MHz. We observe
that, when considering the $X$ and $Y$ axes, the \gls{3dctra} model always
outperforms both the \gls{dr} and the \gls{ctra}+ models, thanks to its richer
representation of the drone's movements. As expected, \gls{ctra}+ also
outperforms \gls{dr}, since it uses the same model as \gls{3dctra} on the
horizontal plane. Surprisingly, when considering the $Z$ axis, the \gls{dr}
system shows a slightly lower error than \gls{ctra}+ and \gls{3dctra}. This
might be due to climbs and dives being relatively rare, so \gls{dr}'s more
frequent updates could give it a slight edge over \gls{3dctra}. On this axis,
\gls{ctra}+ performs worst, because it combines the lower update frequency of
\gls{3dctra} and the inaccurate model of \gls{dr}.

\begin{figure}[t!]
	\centering \setlength{\belowcaptionskip}{-0.5cm}
  \includegraphics[width=0.7\columnwidth,trim={0.5cm 0 1.5cm
    1cm},clip]{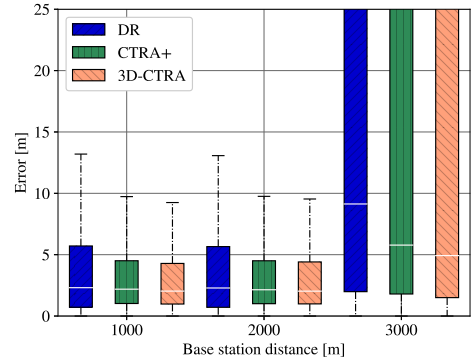}
	\caption{Error for different values of $d$ with SF = 7 and $B=250$ MHz.}
	\label{fig:err_dr6}
\end{figure}

\begin{figure}[t!]
	\centering \setlength{\belowcaptionskip}{-0.5cm}
  \includegraphics[width=0.7\columnwidth,trim={0.5cm 0 1.5cm
    1cm},clip]{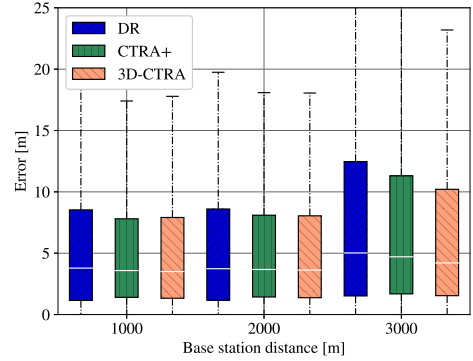}
	\caption{Error for different values of $d$ with SF 8 and $B=250$ MHz.}
	\label{fig:err_dr20}
\end{figure}

We now evaluate how the communication parameters can affect the performance of
the tracking system. Fig.~\ref{fig:err_dr6} shows the boxplot of the positioning
error for different values of $d$ and considering \gls{sf} = 7 and $B=250$ MHz.
It is easy to see that \gls{3dctra} outperforms the other approaches, even if
its updates are less frequent. In particular, considering $d=1000$ m, the 75th
percentile of the error obtained with the \gls{3dctra} model is $30\%$ lower
than \gls{dr}'s and $10\%$ lower than \gls{ctra}+'s. We observe that all the
systems guarantee an average error below $2$ m for both $d=1000$ m and $d=2000$
m. However, the error dramatically increases when considering $d=3000$ m for all
the considered schemes, since the drone is too far away from the control station
for \gls{sf} = 7 and several packets are lost. However, \gls{3dctra} is still
the best option: it is the only one to guarantee an average error below $5$ m.

To allow the control station to accurately predict the drone's trajectory at
larger distances, it is necessary to adopt a more robust communication setting.
In Fig.~\ref{fig:err_dr20}, we report the results we obtained for $d\in\{1000,
2000, 3000\}$, with \gls{sf} = 8 and $B=250$ MHz. Since we increased the
\gls{sf}, the bitrate is lower, and the drone transmits its state less
frequently. As expected, this implies that the positioning error increases at
short distances. For $d=1000$ m and $d=2000$ m, the \gls{3dctra} model
guarantees lower error than the \gls{dr} model but presents similar results to
the \gls{ctra}+ model. Instead, in the case of $d=3000$ m, \gls{3dctra}
outperforms both the other techniques: the 75th percentile error obtained with
the \gls{3dctra} is $20\%$ lower than \gls{dr}'s and $12.5\%$ lower than
\gls{ctra}+'s.

Finally, comparing Fig.~\ref{fig:err_dr6} and Fig.~\ref{fig:err_dr20} shows how
a more robust scheme greatly improves the tracking performance when the distance
between the control station and the \gls{uav} is large. It is worth noting that
\gls{lora} also provides a Data Rate Adaptation mechanism, where the \gls{sf}
employed by the device is set by the controller: in this way, it is possible to
benefit from the increased coverage range achieved with higher \gls{sf} when
necessary, and go back to lower \glspl{sf} when the \gls{uav} is closer to the
\gls{gw} to increase the frequency of transmission messages. Therefore, adapting
the \gls{sf} dynamically will be the best choice in a real scenario, providing
significant performance gains. However, the reactiveness of the adaptive
mechanism should be carefully tuned to avoid instability.

\section{Conclusions and future work}\label{sec:conclusion}

In this work, we presented a tracking system for \glspl{uav}, based on a novel
\gls{3dctra} mobility model and on periodic transmissions over \gls{lora}. Our
system can track a drone's trajectory with high accuracy even when the drone is
at 3 km from the \gls{lora} gateway, and the mobility models we propose
significantly outperform standard \gls{dr}. Moreover, \gls{lora}'s duty
cycle limit is suited to manage swarms of
dozens of drones, as it prevents traffic congestion.

There are several possible avenues of future work: a refinement of the movement
model, including maneuver and mission-level information, might reduce the
tracking error. From the communication side, it would be interesting to
investigate the tracking performance with different data rate adaptation
algorithms, as well as explore features that are not part of the \gls{lora}
standard up to now, like the use of a different frequency plan or of
listen-before-talk instead of applying the duty cycle. Finally, the study of the
behavior of swarms, and possible strategies to avoid packet collision, is
another interesting option that would enable new applications by improving the
tracking accuracy at low cost.

\bibliography{./bibliography} \bibliographystyle{IEEEtran}
\end{document}